\documentclass[aps,pra,reprint,a4paper,superscriptaddress,showkeys,floatfix]{revtex4-2}
\usepackage[utf8]{inputenc}              

\usepackage[english]{babel}              
                                         
\usepackage[T1]{fontenc}                 

\usepackage{lmodern}                     

\usepackage{microtype}                   
\usepackage{xcolor}                      

\usepackage{graphicx}                    
	\graphicspath{{fig/}}                

\usepackage[caption=false]{subfig}
\usepackage{amssymb,amsmath,amsfonts}      

\usepackage{mathtools}
\usepackage{tensor}

\usepackage{braket}                        

\usepackage{tensor}                        

\usepackage[smaller]{cancel}               

\usepackage{upgreek}     
\usepackage{bm}          
\usepackage[colorlinks,citecolor=blue,filecolor=black,linkcolor=black,urlcolor=blue]{hyperref}
\newcommand{\bs}[1]{\boldsymbol{#1}}
\usepackage[pagewise,columnwise]{lineno}
\usepackage[normalem]{ulem}
\usepackage{lipsum}

\begin{document}

\title{On the measurement predictions concerning the intrinsic relativistic spin operator}

\date{\today}

\author{E. R. F. Taillebois}
\email{emile.taillebois@ifgoiano.edu.br}
\affiliation{Instituto Federal Goiano - Campus Avançado Ipameri, 75.780-000, Ipameri, Goiás, Brazil}

\author{A. T. Avelar}
\affiliation{Instituto de Física, Universidade Federal de Goiás, 74.690-900, Goiânia,
Goiás, Brazil}

\begin{abstract}
Although there are several proposals of relativistic spin in the literature, the recognition of intrinsicality as a key characteristic for the definition of this concept is responsible for selecting a single tensor operator that adequately describes such a quantity. This intrinsic definition does not correspond to Wigner's spin operator, which is the definition that is widely adopted in the relativistic quantum information theory literature. Here, the differences between the predictions obtained considering the intrinsic spin and Wigner's spin are investigated. The measurements involving the intrinsic spin are modeled by means of the interaction with an electromagnetic field in a relativistic Stern-Gerlach setup.
\end{abstract}



\maketitle


\section{\label{sec:1}Introduction}

Although the concept of spin has been known for almost a century \cite{Frenkel1926,Thomas1926,Thomas1927}, its adequate relativistic definition remains a subject of intense discussion \cite{Saldanha2012,Saldanha2012a,Palmer2013,Taillebois2013,Caban2013,Caban2013pra,Giacomini2019,Taillebois2021}, a renewed interest on the topic having been motivated mainly by its relevance in the context of relativistic quantum information theory \cite{Czachor1997,Peres2002,Peres2004,Zambianco2019}. The lack of consensus concerning this definition has its roots in ambiguities regarding the properties that such a concept should satisfy \cite{Pryce1948,Fleming1965a,Caban2013pra,Bauke2014,Taillebois2021} and also in its direct relationship with the problem of localization, which suffers from several technical difficulties in the quantum relativistic scenario \cite{Hegerfeldt1974,Hegerfeldt1980,Hegerfeldt1985,Palmer2012,Caban2014,Celeri2016,Silva2019,Taillebois2021a}. However, we have recently shown that the relativistic spin definitions present in literature in general fail to satisfy an essential premise concerning the spin concept: its intrinsicality \cite{Taillebois2021}.

Qualitatively, the definition of an intrinsic property in the special relativity framework can be understood in terms of the dependence of this property with regard to the hyperplane related to which it is defined: the mathematical object that represents an intrinsic property cannot depend on the orientation of the hyperplane on which it is measured (punctual Lorentz invariance) or on the point of intersection of the observer’s world line with the measurement hyperplane (space-like translational invariance). This definition led us in \cite{Taillebois2021} to a unique relativistic spin operator that adequately incorporates the expected non-relativistic behavior together with the fundamental premise of intrinsicality. Adopting Minkowski's metric with signature $(+,-,-,-)$ and Penrose's abstract index notation \cite{Penrose1984}, the intrinsic spin for a system of fixed mass $m$ is given by the tensor operator
\begin{equation*}
\hat{S}^{ab} = \frac{E^{abcd}\hat{P}_{c}\hat{W}_{d}}{(mc)^{2}} = -\frac{i}{\hbar}\frac{[\hat{W}^{a},\hat{W}^{b}]}{(mc)^{2}},
\end{equation*}
with $\hat{W}^{a} \equiv \tensor[^{*}]{\hat{J}}{^{ab}}\hat{P}_{b} = E^{abcd}\hat{J}_{cd}\hat{P}_{b}/2$ standing for the well-known Pauli-Lubanski operator, $E^{abcd}$ the Levi-Civita tensor, $\hat{P}^{b}$ the 4-momentum operator and $\hat{J}^{ab}$ the angular momentum tensor operator.

As any theoretical proposal must be prone to experimental confrontations, it is fundamental to have a measurement model for the intrinsic spin. To this end, a consistent electromagnetic coupling involving the intrinsic spin was presented in \cite{Taillebois2021}, the corresponding interaction Hamiltonian being of the form
\begin{equation}
\hat{\mathcal{H}}_{I} \propto \frac{F^{ab}\hat{S}_{ab}}{2}, \label{eq:Hint1}
\end{equation}
where $F^{ab}$ denotes a classical electromagnetic tensor field. Since \eqref{eq:Hint1} is equivalent to $\hat{\mathcal{H}}_{I} \propto B^{a}(\hat{P})\hat{W}_{a}/(mc)$, with $B^{a}(\hat{P}) = -\tensor[^{*}]{F}{^{ab}}\hat{P}_{b}/(mc)$, this interaction agrees with previous descriptions of relativistic Stern-Gerlach measurements based on the magnetic coupling in the particle's instantaneous rest frame \cite{Saldanha2012,Saldanha2012a,Palmer2013}. This allows to introduce an operationally relevant interpretation for the intrinsic spin tensor, since the electromagnetic-spin interaction can be used to measure specific components of $\hat{S}^{ab}$ even for quantum state in an arbitrary momentum superposition, as opposed to the case of Wigner's spin operator $\hat{\mathbf{S}}$.

Although there is no evidence of any realistic interaction allowing the measurement of specific components of Wigner's operator $\hat{\mathbf{S}}$ for arbitrary quantum states, this operator is widely used in the relativistic quantum information literature as relativistic spin observable. Here, we investigate the differences in the predictions made using this operator and those obtained using the intrinsic spin, which has a well defined measurement procedure associated to it. In order to verify the dependence of the results on the system's velocity, momentum eigenstates of spin-half systems are considered.

The article is organized as follows. In Section \ref{sec:2} we present the eigenvalues and eigenstates associated to the components of the intrinsic spin tensor operator $\hat{S}^{ab}$. In Section \ref{sec:3} the differences between the predictions involving Wigner's spin operator $\hat{\mathbf{S}}$ and the intrinsic spin operator $\hat{S}^{ab}$ are investigated for different scenarios. Finally, concluding remarks are presented in Section \ref{sec:4}. The framework of irreducible unitary representations (IURs) of the Poincaré group \cite{Wigner1939,Tung1985} is adopted in all sections, a thorough explanation of its connection to Dirac spinor formalism being found in \cite{Caban2005}. 

\section{\label{sec:2}Intrinsic spin eigenstates}

Since $\hat{S}^{ab}$ is an antisymmetric tensor, it can always be written as
\begin{equation*}
\hat{S}^{ab} = 2\hat{\mathcal{V}}^{[a}(v)v^{b]} + \frac{1}{c}E^{abcd}v_{c}\hat{\Sigma}_{d}(v),
\end{equation*}
where $\hat{\Sigma}^{a}(v) = \tensor[^{*}]{\hat{S}}{^{ab}}v_{b}/c$ and $\hat{\mathcal{V}}^{a}(v) = \hat{S}^{ab}v_{b}/c^{2}$ represent, respectively, the angular momentum and the mass momentum components of the intrinsic tensor $\hat{S}^{ab}$ with regard to an inertial observer of 4-velocity $v^{a}$. Using an orthonormal basis $\{e^{a}_{(\mu)}\}$ such that $v^{0} = c$, as well as the notation $\hat{\mathbf{S}}_{\parallel} = (\hat{\mathbf{S}}\cdot\hat{\mathbf{P}})\hat{\mathbf{P}}/\|\hat{\mathbf{P}}\|^{2}$ and $\hat{\mathbf{S}}_{\perp} = \hat{\mathbf{S}} - \hat{\mathbf{S}}_{\parallel}$, these 4-vector quantities can be written as $\hat{\Sigma}^{a}(v) \dot{=} (0,\hat{\bs{\Sigma}}(v))$ and $\hat{\mathcal{V}}^{a}(v) \dot{=} (0,\hat{\bs{\mathcal{V}}}(v))$, with
\begin{subequations}\label{eq:Mom}
\begin{eqnarray}
\hat{\bs{\Sigma}}(v) & = & (\hat{S}^{23},\hat{S}^{31},\hat{S}^{12}) = \frac{\hat{P}^{0}}{mc}\hat{\mathbf{S}}_{\perp} + \hat{\mathbf{S}}_{\parallel}, \label{eq:AngMom}\\
\hat{\bs{\mathcal{V}}}(v) & = & \frac{1}{c}(\hat{S}^{10},\hat{S}^{20},\hat{S}^{30}) = \frac{\hat{\mathbf{P}}\times\hat{\mathbf{S}}}{mc^2}. \label{eq:MasMom} 
\end{eqnarray}
\end{subequations}

In what follows, the eigenvalues and eigenstates of such operators are obtained using the formalism of IURs of the Poincaré group. In this framework, single-particle base states of fixed mass $m$ and spin $s$ are labeled by their momentum $\mathbf{p}$ and a secondary spin number associated to an operator that depends on the complementary set used to define the basis \cite{Taillebois2013}. Adopting the boost complementary set, this operator coincides with Wigner's spin operator $\hat{\mathbf{S}}$ projected in an arbitrary fixed direction. Thus, choosing the $i$-axis as reference, these base states are written as $\ket{\mathbf{p},m^{i}}_{B}$, with $\hat{S}^{i}\ket{\mathbf{p},m^{i}}_{B} = \hbar m^{i}\ket{\mathbf{p},m^{i}}_{B}$ and $m^{i} = -s, -s+1, \dots, s-1, s$, the index $B$ indicating the choice of the boost complementary set.  

\subsection{\label{subsec:EE} Eigenvalues and eigenstates of \texorpdfstring{$\hat{\Sigma}^{i}(v)$}{S12}}

Taking into account that $[\hat{P}^{\alpha}, \hat{S}^{\mu\nu}] = 0$, a basis of momentum eigenstates can be constructed by choosing one of the components of the operator $\hat{\bs{\Sigma}}(v)$ to form a complete set of commuting observables (CSCO).  Labeling such eigenstates as $\ket{\mathbf{p}, \alpha^{i}}_{\Sigma}$, with $\hat{\Sigma}^{i}(v)\ket{\mathbf{p}, \alpha^{i}}_{\Sigma} = \hbar\alpha^{i}\ket{\mathbf{p}, \alpha^{i}}_{\Sigma}$, it follows from \eqref{eq:AngMom} that the eigenvalues of $\hat{\Sigma}^{i}(v)$ are given by $\alpha^{i} = \pm s_{p}^{i}$, with $s_{p}^{i} \equiv \tfrac{1}{2}\sqrt{1 + (\tilde{p}^j)^2 + (\tilde{p}^k)^2}$, the tilde notation representing the dimensionless quantities $\tilde{\mathbf{p}} = \mathbf{p}/mc$. The corresponding eigenstates are given by
\begin{equation*}
\ket{\mathbf{p},\varepsilon s_{p}^{i}}_{\Sigma} = a^{(+)}_{p,i}\ket{\mathbf{p},m^{i} = \tfrac{\varepsilon}{2}}_{B} + \varepsilon e^{i\varepsilon\varphi_{p}^{i}}a^{(-)}_{p,i}\ket{\mathbf{p},m^{i} = -\tfrac{\varepsilon}{2}}_{B},
\end{equation*}
where
\begin{equation*}
a^{(\xi)}_{p,i} = \frac{1}{2}\sqrt{\frac{(\tilde{E}_{p} + 2\xi s_{p}^{i})(2s_{p}^{i} + \xi)}{s_{p}^{i}(1 + \tilde{E}_{p})}},
\end{equation*}
with $\tilde{E}_{p} = \sqrt{1 + \|\mathbf{\tilde{p}}\|^{2}}$, and
\begin{equation*}
\begin{aligned}
e^{i\varphi_{p}^{3}} & = -\frac{\mathrm{sgn}(\tilde{p}^{3})(\tilde{p}^{1} + i\tilde{p}^{2})}{\sqrt{(\tilde{p}^1)^2 + (\tilde{p}^2)^2}}, \\
e^{i\varphi_{p}^{2}} & = -\frac{\mathrm{sgn}(\tilde{p}^{2})(\tilde{p}^{3} + i\tilde{p}^{1})}{\sqrt{(\tilde{p}^3)^2 + (\tilde{p}^1)^2}}, \\
e^{i\varphi_{p}^{1}} & = -\frac{\mathrm{sgn}(\tilde{p}^{1})(\tilde{p}^{3} - i\tilde{p}^{2})}{\sqrt{(\tilde{p}^2)^2 + (\tilde{p}^3)^2}}.
\end{aligned}
\end{equation*}

It is noteworthy that the spectrum of $\hat{\Sigma}^{i}(v)$ is momentum dependent. More specifically, the eigenvalues of $\hat{\Sigma}^{i}(v)$ depend on the norm of the momentum component perpendicular to the $i-$axis direction. Despite this dependence, for each momentum the spectrum continues to assume only two possible values. This result only highlights the fiber bundle structure of the Hilbert space associated with the description of a massive particle, as presented in \cite{Taillebois2013,Palmer2013}. It is worth remembering that, although there is no explicit momentum index in the Wigner spin term of $\ket{\mathbf{p},m^{i}}_{B}$, the aforementioned fiber bundle structure leads to an implicit momentum dependence in the secondary spin number $m^{i}$ \cite{Taillebois2013}. 

\subsection{Eigenvalues and eigenstates of \texorpdfstring{$\hat{\mathcal{V}}^{3}(v)$}{S30}}

From $[\hat{P}^{\alpha}, \hat{S}^{\mu\nu}] = 0$ it follows that a basis of momentum eigenstates can also be constructed by choosing one of the components of $\bs{\hat{\mathcal{V}}}(v)$ to form a CSCO. Labeling such eigenstates as $\ket{\mathbf{p}, \beta^{i}}_{\mathcal{V}}$, with $\hat{\mathcal{V}}^{i}(v)\ket{\mathbf{p}, \beta^{i}}_{\mathcal{V}} = \tfrac{\hbar}{c}\beta^{i}\ket{\mathbf{p}, \beta^{i}}_{\mathcal{V}}$, it follows from \eqref{eq:MasMom} that the eigenvalues of $\hat{\mathcal{V}}^{i}(v)$ are given by $\beta^{i} = \pm \mu_{p}^{i}$, with $\mu_{p}^{i} \equiv \tfrac{1}{2}\sqrt{(\tilde{p}^j)^2 + (\tilde{p}^k)^2}$. Thus, as for $\hat{\Sigma}^{i}(v)$, the eigenvalues of $\hat{\mathcal{V}}^{i}(v)$ are momentum dependent, the dependence being associated to the norm of the momentum perpendicular to the $i-$axis direction. For $\sqrt{(\tilde{p}^j)^2 + (\tilde{p}^k)^2} \neq 0$, the associated eigenstates can be written as
\begin{equation*}
\ket{\mathbf{p},\varepsilon\mu_{p}}_{\mathcal{V}} = \frac{1}{\sqrt{2}}\left(\ket{\mathbf{p},m^{i}=-\tfrac{\varepsilon}{2}}_{B} + \varepsilon e^{i\varepsilon\phi^{i}_{p}}\ket{\mathbf{p},m^{i}=\tfrac{\varepsilon}{2}}_{B}\right),
\end{equation*}
where
\begin{equation*}
\begin{aligned}
e^{i\phi^{3}_{p}} & = -\frac{(\tilde{p}^{2} + i\tilde{p}^{1})}{\sqrt{(\tilde{p}^1)^2 + (\tilde{p}^2)^2}}, \\
e^{i\phi^{2}_{p}} & = -\frac{(\tilde{p}^{1} + i\tilde{p}^{3})}{\sqrt{(\tilde{p}^3)^2 + (\tilde{p}^1)^2}}, \\
e^{i\phi^{1}_{p}} & = -\frac{(-\tilde{p}^{2} + i\tilde{p}^{3})}{\sqrt{(\tilde{p}^2)^2 + (\tilde{p}^3)^2}}.
\end{aligned}
\end{equation*}
For $\sqrt{(\tilde{p}^j)^2 + (\tilde{p}^k)^2} = 0$ the eigenvalues of $\hat{\mathcal{V}}^{i}(v)$ are degenerate and the corresponding eigenstates are simply given by $(\ket{(0,0,p^i),m^{i}=\tfrac{1}{2}}_{B} \pm \ket{(0,0,p^i),m^{i} = -\tfrac{1}{2}}_{B})/\sqrt{2}$.

\section{\label{sec:3}Discrepancies between Wigner's spin and the intrinsic spin}

\subsection{Measurement setup}

Given a classical electromagnetic tensor field $F^{ab}$, the electric and magnetic field as seen by an observer of 4-velocity $v^{a}$ can be described by the 4-vector quantities $E^{a}(v) = F^{ab}v_{b}$ and $B^{a}(v) = - \tensor[^{*}]{F}{^{ab}}v_{b}/c$. Using an orthonormal basis $\{e^{a}_{(\mu)}\}$ such that $v^{0} = c$, these 4-vector quantities can be written as $E^{a}(v) \dot{=} (0,\mathbf{E}(v))$ and $B^{a}(v) \dot{=} (0,\mathbf{B}(v))$, where $\mathbf{E}(v)$ and $\mathbf{B}(v)$ are, respectively, the electric and the magnetic vector fields as seen from the laboratory frame. Thus, using the notation in \eqref{eq:Mom}, the interaction Hamiltonian in \eqref{eq:Hint1} can be rewritten as
\begin{equation}
\hat{\mathcal{H}}_{I} \propto -\left(\mathbf{B}(v)\cdot\hat{\bs{\Sigma}}(v) + \mathbf{E}(v)\cdot\hat{\bs{\mathcal{V}}}(v)\right),
\label{eq:Hint3}
\end{equation}
which explicits the fact that the components of the intrinsic spin tensor can be selected using classical electromagnetic fields in the laboratory frame, even for quantum states given by an arbitrary momentum superposition. From \eqref{eq:Hint3}, the angular momentum contribution of the intrinsic tensor spin is associated to measurements involving magnetic fields, while the momentum mass contribution is associated to measurements involving electric fields. It is worth noting that the interaction Hamiltonian \eqref{eq:Hint3} leads to twice the expected electric interaction, since the Thomas correction \cite{Thomas1927} for the contribution of the electron acceleration was not taken into account.

In what follows, different setups will be considered to confront the predictions obtained using Wigner's spin with those resulting from the use of the intrinsic spin operator. Two different input states will be considered for each setup, both being momentum eingenstates, one in Wigner's basis $\{\ket{\mathbf{p},m^{i}}_{B}\}$ and the other in the intrinsic spin basis $\{\ket{\mathbf{p},\alpha^{i}}_{\Sigma}\}$. The first choice of basis allows to assess the influence of the Stern-Gerlach measurements on the results obtained considering the base states usually adopted in the literature, while the second choice allows to investigate the consequences of considering this type of measurement both in the state creation process and in the final measurement.

Since the complementary set that defines the Wigner states is composed only by boosts, a state of the form $\ket{\mathbf{p},m^{i} = \tfrac{1}{2}}_{B}$ can be easily conceived by assuming that a particle created with a spin up component relative to its own rest frame $i$-axis is to be measured by an observer with a 4-velocity $v^{a} = \mathbb{P}p^{a}/m$, where  $m$ is the mass of the particle, $\mathbb{P}$ indicates a parity transformation and $p^{a}$ is a 4-vector quantity such that $p^{a} \dot{=} (E_{p}/c,\mathbf{p})$ in the rest frame of the particle. On the other hand, the states $\ket{\mathbf{p},\alpha^{i}}_{\Sigma}$ can be conceived by assuming a first relativistic Stern-Gerlach measurement with a magnetic field aligned with the $i$-axis of the laboratory frame and a momentum selector filter.

\subsection{Interaction with a magnetic field}

In this section we will assume that the observer performs spin measurements using a magnetic field aligned with the laboratory frame z-axis, i.e., $B^{a}(v) \dot{=} (0,0,0,B^{3}(v))$. From \eqref{eq:Hint3}, this corresponds simply to measuring the component $\hat{\Sigma}^{3}(v)$ of the intrinsic spin operator by using a Stern-Gerlach apparatus.

\subsubsection{Magnetic field parallel to the input spin}

\begin{figure}[b]
\centering
\includegraphics[width=0.85\linewidth]{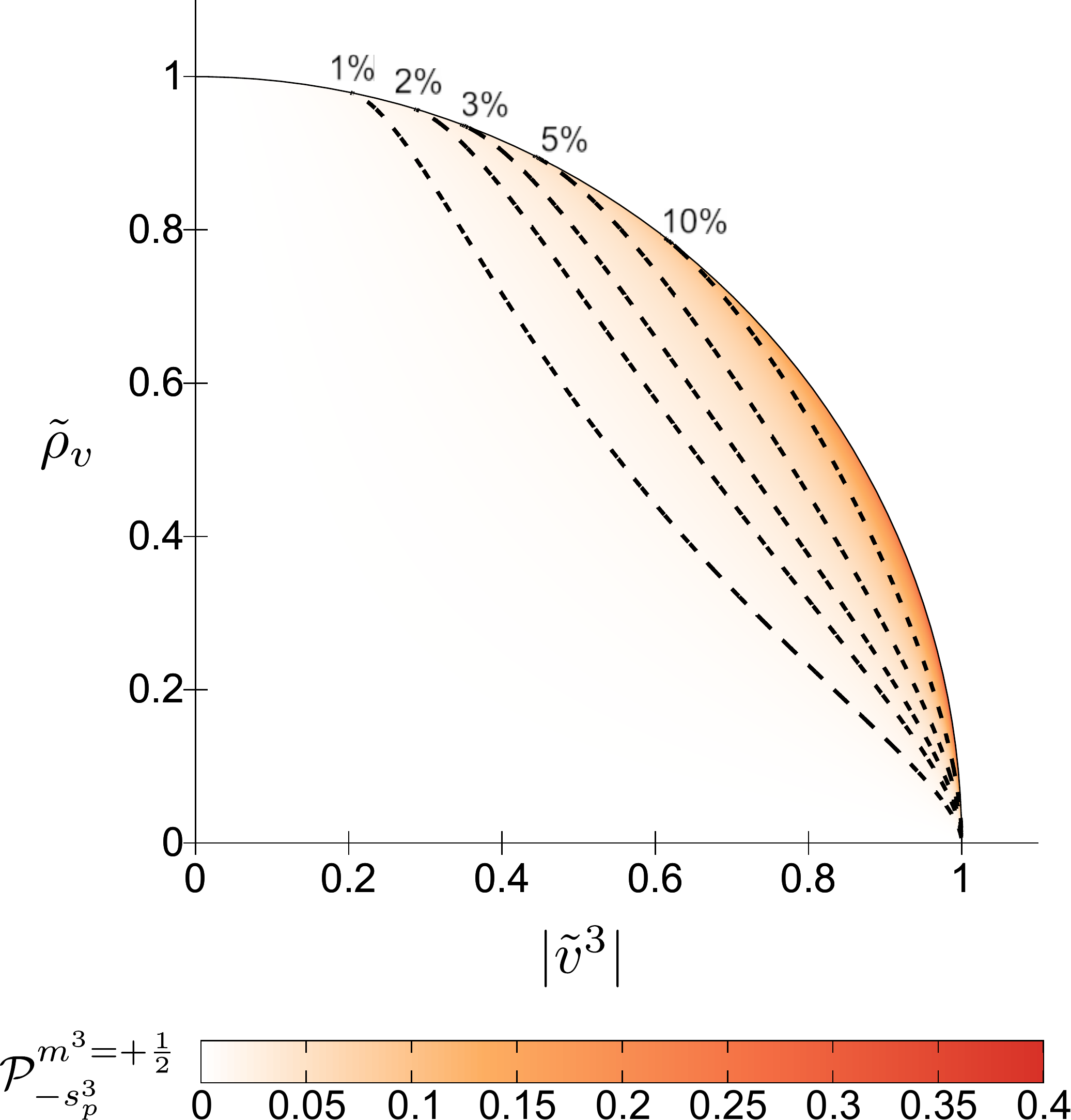}
\caption{\label{fig:ProbV3Vrho} Probability of finding the negative eigenvalue $-s^{3}_{p}$ for an input state given by a Wigner momentum eigenstate with $m^{3} = \tfrac{1}{2}$. For a measurement performed using the intrinsic spin definition, which corresponds to a relativistic Stern-Gerlach setup, there is a probability of finding such negative outputs, in contrast with the case of a hypothetical measurement of $\hat{S}^{3}$. The dotted lines represent a set of some lower bound values of $\Delta$ for the probability to be considered significant.}
\end{figure}

First, we will assume that the system to be measured is created in a initial state given by $\ket{\mathbf{p},m^{3} = \tfrac{1}{2}}_{B}$. With such an input state, if the observer had an hypothetical apparatus capable of measuring Wigner's spin operator, all the results should indicate the positive eigenvalue $+\hbar/2$ for a measurement of $\hat{S}^{3}$. However, if the available apparatus relies on the electromagnetic interaction, as in a relativistic Stern-Gerlach experiment, discrepancies with respect to those predictions will be verified. Thus, in contrast to the prediction concerning  Wigner's spin, it follows that it is possible for the observer to find the negative eigenvalue $-s_{p}^{3}$, the associate probability being given by
\begin{equation}
\mathcal{P}_{-s_{p}^{3}}^{m^{3}=+\frac{1}{2}} = \frac{1}{4}\frac{(\tilde{E}_{p} - 2s_{p}^{3})(2s_{p}^{3} - 1)}{s_{p}^{3}(1+\tilde{E}_{p})}. \label{eq:ProbZ}
\end{equation} 

From an experimental perspective, it is interesting to evaluate for which particle's velocities the probability \eqref{eq:ProbZ} is in fact significant. Setting $\mathbf{\tilde{p}} = \gamma\mathbf{\tilde{v}}$ in \eqref{eq:ProbZ}, with $\mathbf{\tilde{v}} = \mathbf{v}/c$ being the dimensionless velocity and $\gamma = (1 - \|\mathbf{\tilde{v}}\|^{2})^{-1/2}$ the usual Lorentz factor, this probability can be plotted as a map of the variables $\tilde{v}^{3}$ and $\tilde{\rho}_{v} = \sqrt{(\tilde{v}^{1})^{2} + (\tilde{v}^{2})^{2}}$ (Fig. \ref{fig:ProbV3Vrho}). From this map, the relevance of the negative outcomes for the proposed measurement can be evaluated by choosing hypothetical lower bound values $\Delta$ for the probability to be considered significant. In Fig. \ref{fig:ProbV3Vrho}, a set of some significance boundaries are represented as dotted lines, the respective lower bound values $\Delta$ being indicated at the top of the circular map of velocities.

The map in Fig. \ref{fig:ProbV3Vrho} makes evident that, although the probability \eqref{eq:ProbZ} is in general non-null, it is small (less than $1\%$) for a great set of velocities, even relativistic ones. Besides that, it makes evident the relevance of the velocity component in the direction of the magnetic field for the negative outcomes to be significant: each value of $\Delta$ implies a relevant lower bound for $|\tilde{v}^{3}|$, the velocity component parallel to the magnetic field. The fact that a significant amount of velocity is needed in the direction parallel to the field is relevant for the conception of measurements based on a relativistic Stern-Gerlach apparatus, since this may lead to difficulties concerning the dimensions of the equipment.

\begin{figure}[t]
\includegraphics[width=\linewidth]{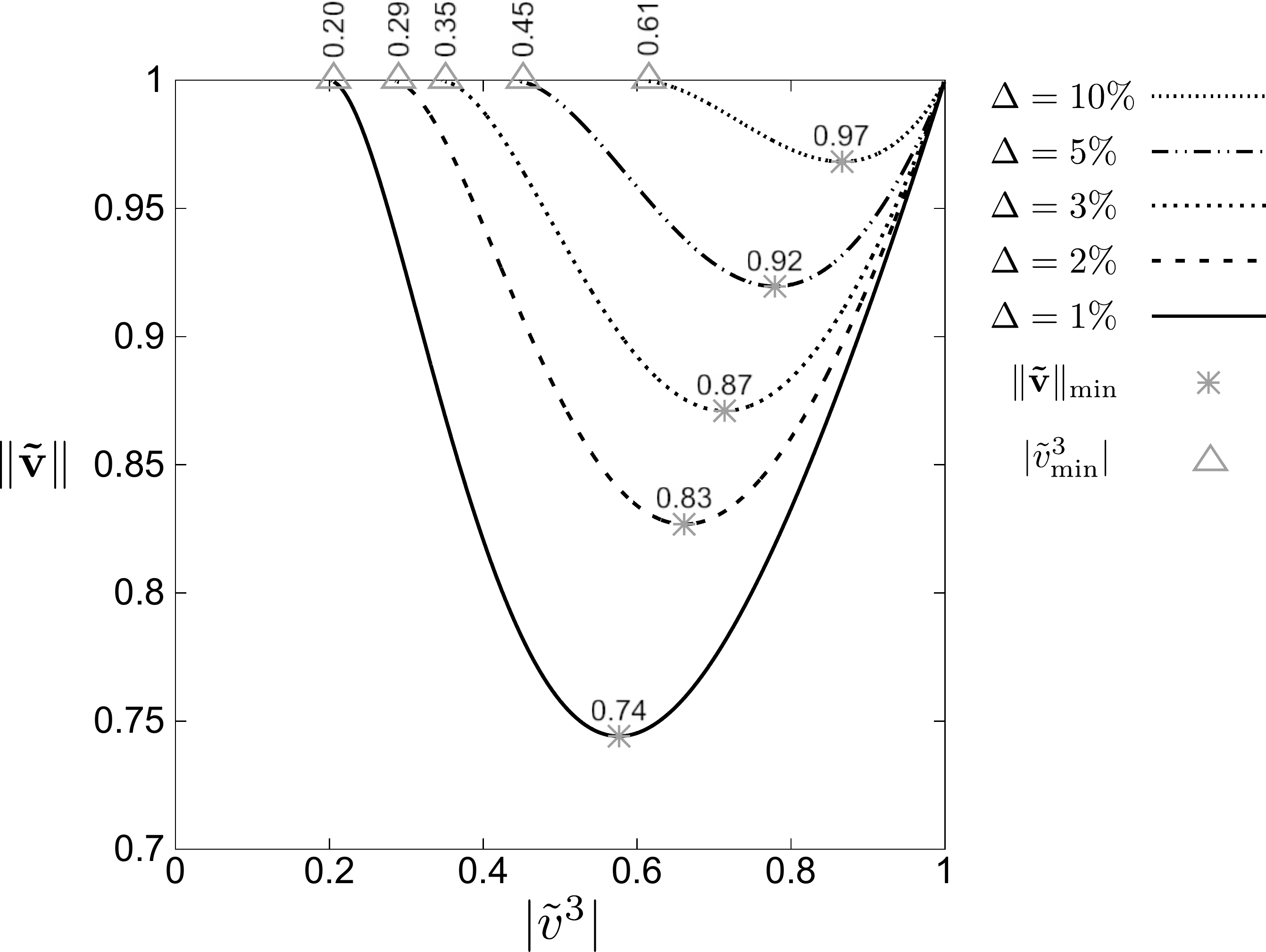}
\caption{\label{fig:Vmod} Plot of the dimensionless velocity norm as a function of $|\tilde{v}^{3}|$ for several values of the significance window $\Delta$. For each value of $\Delta$ the minimum value of $\|\mathbf{\tilde{v}}\|$ as well as the minimum value of $|\tilde{v}^{3}|$ are indicated.}
\end{figure}

To evaluate the minimum value of $|\tilde{v}^{3}|$ and the amount of relativistic velocity needed to observe the predicted discrepancies, a plot of the dimensionless velocity norm $\|\mathbf{\tilde{v}}\|$ as a function of $|\tilde{v}^{3}|$ is presented in Fig. \ref{fig:Vmod} for several values of significance window $\Delta$. For each value of $\Delta$ the minimum value of $\|\mathbf{\tilde{v}}\|$ as well as the minimum value of $|\tilde{v}^{3}|$ are indicated. This plot shows that, for each significance window $\Delta$, the minimum amount of velocity norm does not correspond to the minimum of $\tilde{v}^{3}$. Thus, to avoid great values of $\tilde{v}^{3}$, which would imply technical difficulties concerning the dimensions of the equipment, the observation of the phenomenon would require greater relativistic velocities, which also introduce challenging difficulties.

For this setup, with a magnetic field parallel to the spin sate, the case of an intrinsic spin input state of the form $\ket{\mathbf{p},+s^{3}_{p}}_{B}$ will not be considered, since the proposed measurement will return $100\%$ of the results with spin up for all values of particle momentum.

\subsubsection{Magnetic field perpendicular to the input spin}

In this second setup, $\ket{\mathbf{p},m^{1} = \tfrac{1}{2}}_{B}$ will be chosen as first input state. Thus, an hypothetical measurement of Wigner's spin operator $\hat{S}^{3}$ would lead to both negative and positive eigenvalues with equal probabilities. However, assuming that the observer performs the measurements using a Stern-Gerlach apparatus with a magnetic field aligned to the laboratory z-axis, discrepancies relative to the Wigner spin predictions will emerge.

To evaluate the aforementioned discrepancies we will look for the probability difference $d_{+s_{p}^{3}}^{m^{1} = +\frac{1}{2}} \equiv \mathcal{P}_{+s_{p}^{3}}^{m^{1} = +\frac{1}{2}} - \tfrac{1}{2}$, where $\mathcal{P}_{+s_{p}^{3}}^{m^{1} = +\frac{1}{2}}$ is the probability of finding the positive eigenvalue $+s^{3}_{p}$ with a Stern-Gerlach measurement and $\tfrac{1}{2}$ represent the probability of obtaining a positive eigenvalue for a measurement of Wigner's spin $\hat{S}^{3}$. Using the results presented in Section \ref{subsec:EE}, this difference can be shown to be given by
\begin{equation}
d_{+s_{p}^{3}}^{m^{1} = +\frac{1}{2}} = -\frac{1}{2}\frac{\tilde{v}^{3}\tilde{v}^{1}}{(\gamma^{-1} + 1)\sqrt{1 - (\tilde{v}^{3})^{2}}}.
\label{eq:DifProb1}
\end{equation}

\begin{figure}[t]
\includegraphics[width=\linewidth]{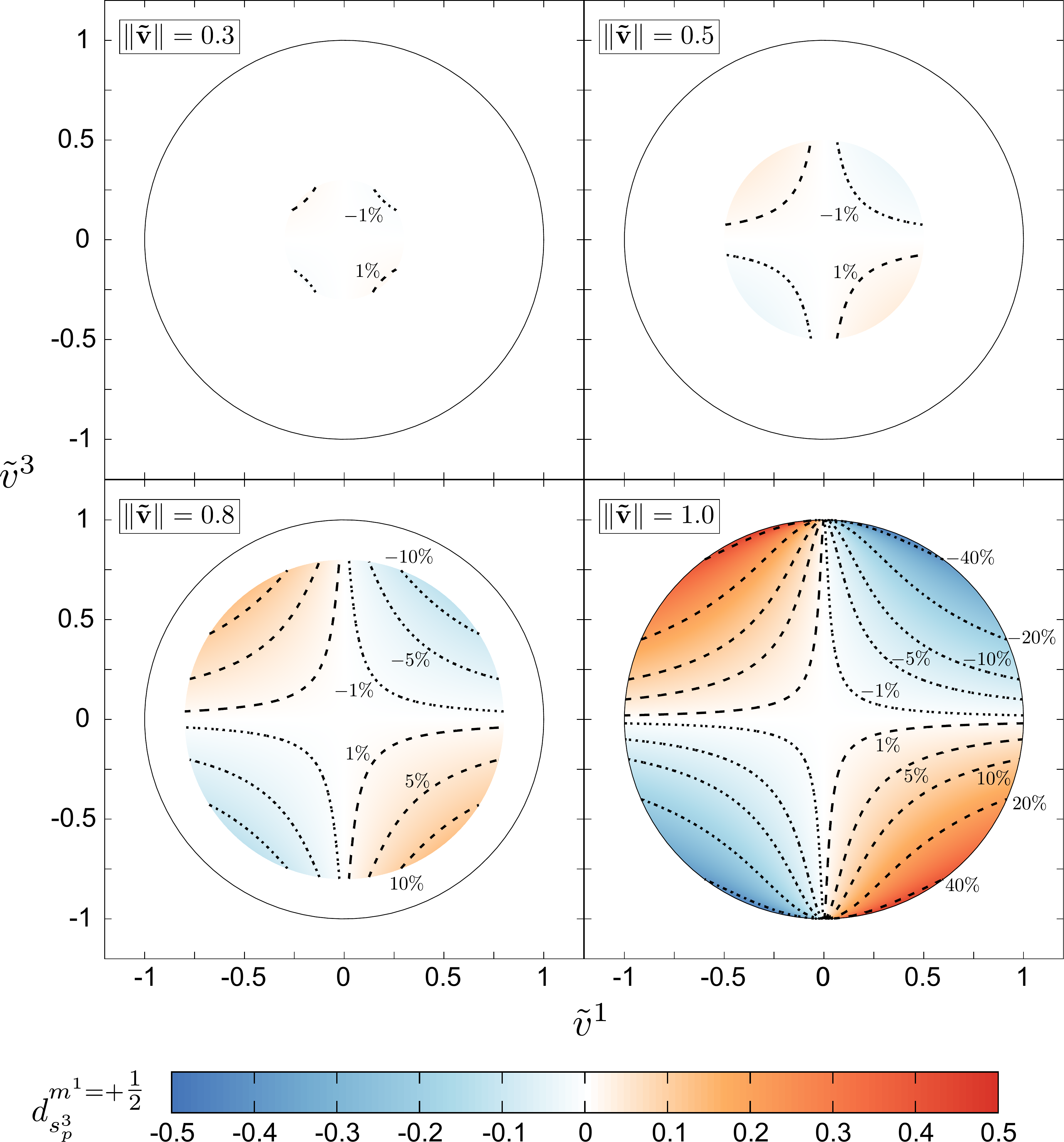}
\caption{\label{fig:ProbV1V3B1SG3} Probability difference $d_{+s_{p}^{3}}^{m^{1} = +\frac{1}{2}}$ in terms of $\tilde{v}^{1}$ and $\tilde{v}^{3}$ for several fixed values of $\|\mathbf{\tilde{v}}\|$. The value of $\tilde{v}^{2}$ grows radially as we approach the center of the velocity circles. Positive (negative) contour lines are presented as dashed (dotted) lines for a better visualization of the velocity dependence.}
\end{figure}

In Fig. \ref{fig:ProbV1V3B1SG3}, the probability difference \eqref{eq:DifProb1} is plotted in terms of $\tilde{v}^{1}$ and $\tilde{v}^{3}$ for several fixed values of $\|\mathbf{\tilde{v}}\|$. This figure shows that the differences with the previsions made using Wigner's spin operator are most easily observed in the case of perpendicular-to-field input spin state, discrepancy values of $10\%$ being already observable for $\|\mathbf{\tilde{v}}\|$ in the order of $0.8$. However, as for the case of a magnetic field parallel to a Wigner spin input state, Fig. \ref{fig:ProbV1V3B1SG3} shows that the velocity component $\tilde{v}^{3}$ parallel to the direction of the magnetic field also plays a fundamental role in the observation of significant values of $d_{+s_{p}^{3}}^{m^{1}=+\tfrac{1}{2}}$, greater values of this quantity being achievable only with considerable contributions of the velocity in the field direction. Thus, if great values of $|\tilde{v}^{3}|$ are to be avoided, discrepancies will be significantly observable only at the expense of greater values of $|\tilde{v}^{1}|$. It is worth noting that, for fixed values of $\|\mathbf{\tilde{v}}\|$, larger values of $|\tilde{v}^{2}|$ do not increase the observability of the discrepancy $d_{+s_{p}^{3}}^{m^{1}=+\tfrac{1}{2}}$, in contrast to what is the case for $|\tilde{v}^{1}|$.

\begin{figure}[t]
\includegraphics[width=\linewidth]{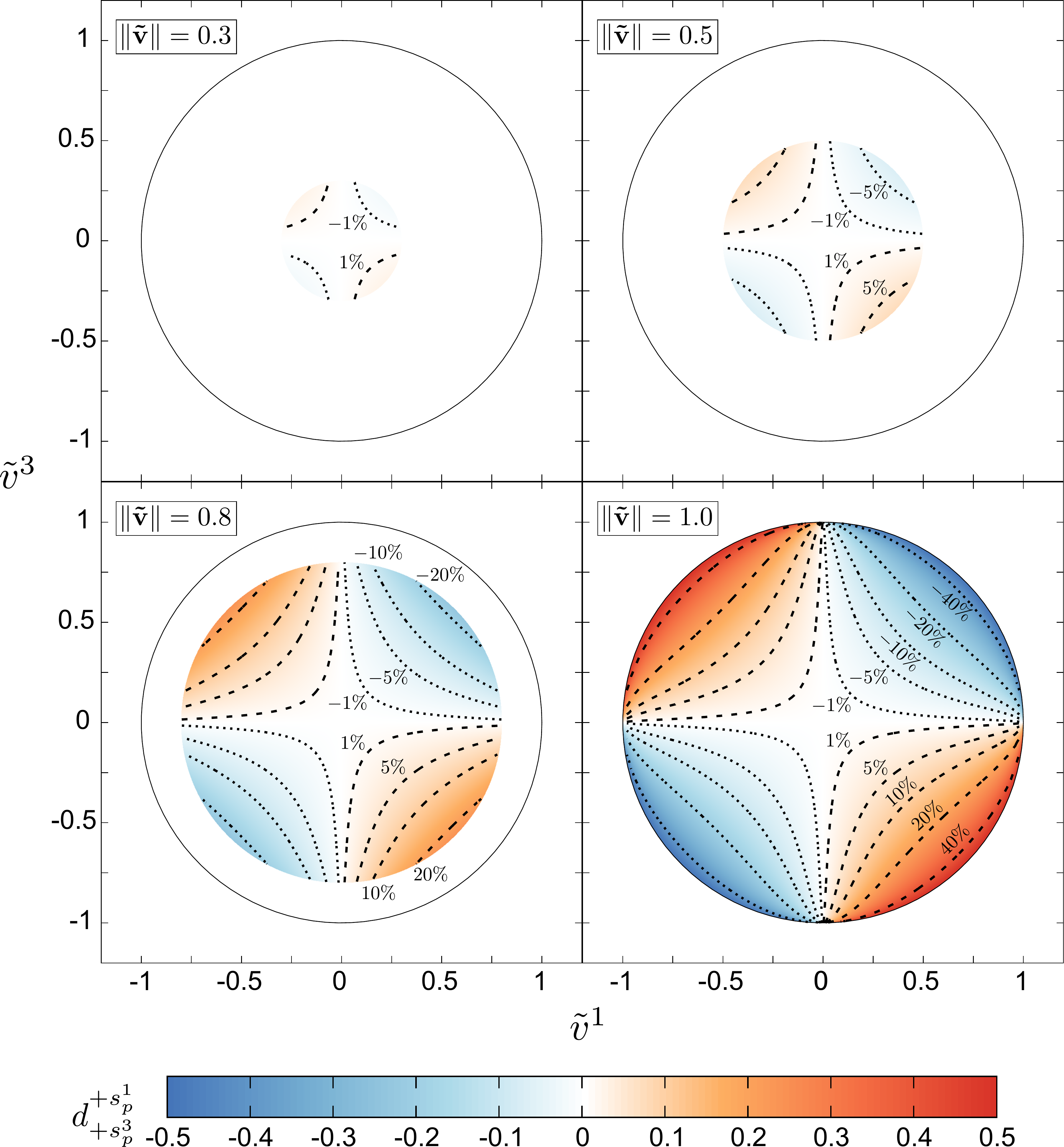}
\caption{\label{fig:ProbV1V3SG1SG3} Probability difference $d_{+s_{p}^{3}}^{+s^{1}_{p}}$ in terms of $\tilde{v}^{1}$ and $\tilde{v}^{3}$ for several fixed values of $\|\mathbf{\tilde{v}}\|$. The value of $\tilde{v}^{2}$ grows radially as we approach the center of the velocity circles. Positive (negative) contour lines are presented as dashed (dotted) lines for a better visualization of the velocity dependence. Due to the form of \eqref{eq:DifProb2}, maps for $\|\mathbf{\tilde{v}}\|<1$ are simply cropped copies of the map for $\|\mathbf{\tilde{v}}\|=1$.}
\end{figure}

Now we can observe the consequences of assuming a state $\ket{\mathbf{p},s^{1}_{p}}_{\Sigma}$ as input state, i.e. a state produced by a procedure using a relativistic Stern-Gerlach apparatus with a magnetic field in the x-axis direction. In this case, the probability discrepancy $d_{+s_{p}^{3}}^{+s^{1}_{p}} \equiv \mathcal{P}_{+s_{p}^{3}}^{+s^{1}_{p}} - \tfrac{1}{2}$, where $\mathcal{P}_{+s_{p}^{3}}^{+s^{1}_{p}}$ is the probability of finding the positive eigenvalue $+s^{3}_{p}$ with a Stern-Gerlach measurement, is given by
\begin{equation}
d_{+s_{p}^{3}}^{+s^{1}_{p}} = -\frac{1}{2}\frac{\tilde{v}^{3}\tilde{v}^{1}}{\sqrt{1 - (\tilde{v}^{1})^{2}}\sqrt{1 - (\tilde{v}^{3})^{2}}}.
\label{eq:DifProb2}
\end{equation}

In Fig. \ref{fig:ProbV1V3SG1SG3}, the probability difference \eqref{eq:DifProb2} is plotted in terms of $\tilde{v}^{1}$ and $\tilde{v}^{3}$ for several fixed values of $\|\mathbf{\tilde{v}}\|$. This figure shows that the probability difference is more easily observed in this setup, since greater values of $d_{+s_{p}^{3}}^{+s^{1}_{p}}$ can be achieved for smaller values of $\|\mathbf{\tilde{v}}\|$ when compared to the case presented in Fig. \ref{fig:ProbV1V3B1SG3}. We observe that discrepancies of the order of $5\%$ are already observable for $\|\mathbf{\tilde{v}}\| = 0.5$ and, therefore, ultrarelativistic velocities are not necessary for the observation of discrepancies with respect to predictions involving the Wigner spin.

The maps in Fig. \ref{fig:ProbV1V3SG1SG3} also show that, in contrast with the previous cases, the velocity component parallel to the measurement field plays a smaller role for the probability difference $d_{+s_{p}^{3}}^{+s^{1}_{p}}$. However, since in this case we assume that the input state was produced by a Stern-Gerlach setup aligned to the x-axis direction, greater values of $\tilde{v}^{1}$ are also to be avoided, a restriction that contributes to reduce the visibility of the discrepancies.

\subsection{Interaction with an electric field}

For completeness, it is interesting to evaluate the predictions that results from assuming an electric field interaction with the intrinsical spin. It is important to note that this interaction arises simply from the fact that an electric field in the laboratory frame implies the existence of a magnetic field in the instantaneous rest frame of the particle. In this third setup, we will assume that the interaction is performed using an electric field aligned to the laboratory frame z-axis, i.e., $E^{a}(v) \dot{=} (0,0,0,E^{3}(v))$. From \eqref{eq:Hint3}, this corresponds to measuring the component $\hat{\mathcal{V}}^{3}(v)$ of the intrinsic spin operator. Since there is not an analogous of $\bs{\hat{\mathcal{V}}}(v)$ in the context of Wigner's spin, the predictions concerning the electrical interaction will not be compared with some hypothetical prediction resulting from Wigner's spin. However, it is interesting to evaluate the differences that arise when assuming that the input state is a Wigner spin eigenstate or a intrinsical spin eigenstate.

\begin{figure}[t]
\centering
\includegraphics[width=0.85\linewidth]{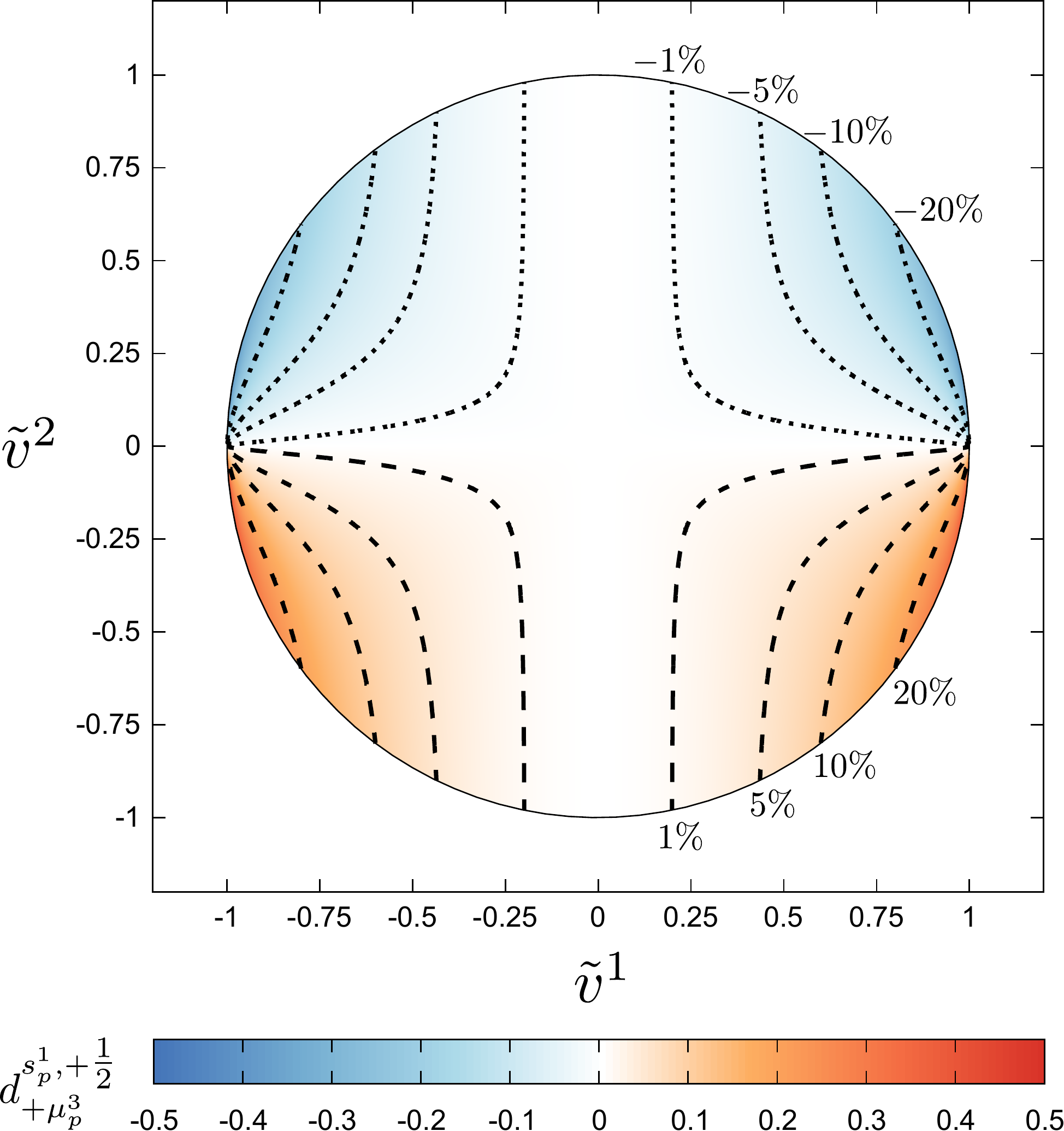}
\caption{\label{fig:ProbV1V2} Probability difference $d^{s^{1}_{p},+\tfrac{1}{2}}_{+\mu_{p}^{3}}$ in terms of $\tilde{v}^{1}$ and $\tilde{v}^{2}$. The maximum value of $\tilde{v}^{3}$ grows radially as we approach the center of the velocity circle. Positive (negative) contour lines are presented as dashed (dotted) lines for a better visualization of the velocity dependence.}
\end{figure}

Assuming $\ket{\mathbf{p},m^{3} = \tfrac{1}{2}}_{B}$ as input state, it follows that there are equal probabilities of finding both signs of the eigenvalues of $\hat{\mathcal{V}}^{3}(v)$ for any non-zero value of the particle's momentum. However, the visibility of this separation will decrease for non-relativistic velocities, since both the positive and negative eigenvalues approach zero as the particle velocity decreases. Now, assuming $\ket{\mathbf{p},m^{1} = \tfrac{1}{2}}_{B}$ as input state, differences in the probability values of the $\hat{\mathcal{V}}^{3}(v)$ measurements will be observed for different values of velocity input. With such an input state, the probability of finding a positive eigenvalue of $\hat{\mathcal{V}}^{3}(v)$ will be given by
\begin{equation}
\mathcal{P}_{+\mu_{p}^{3}}^{m^{1} = +\frac{1}{2}} = \frac{1}{2}\left(1 - \frac{\tilde{v}^{2}}{\tilde{\rho}_{v}}\right)
\end{equation}
On the other hand, if a state $\ket{\mathbf{p},s^{1}_{p}}_{\Sigma}$ is chosen as input state, the probability of finding a positive eigenvalue of $\hat{\mathcal{V}}^{3}(v)$ will be given by
\begin{equation}
\mathcal{P}_{+\mu_{p}^{3}}^{+s^{1}_{p}} = \frac{1}{2}\left(1 - \frac{\tilde{v}^{2}}{\tilde{\rho}_{v}\sqrt{1-(\tilde{v}^{1})^{2}}}\right)
\end{equation}
Since those results do not depend on $\tilde{v}^{3}$, the probability difference $d^{s^{1}_{p},+\tfrac{1}{2}}_{+\mu_{p}^{3}} = \mathcal{P}_{+\mu_{p}^{3}}^{+s^{1}_{p}} - \mathcal{P}_{+\mu_{p}^{3}}^{m^{1} = +\tfrac{1}{2}}$ is plotted in Fig. \ref{fig:ProbV1V2} in terms of $\tilde{v}^{1}$ and $\tilde{v}^{2}$.

The map in Fig. \ref{fig:ProbV1V2} shows that, for the electric interaction, the differences in considering input states $\ket{\mathbf{p},m^{1} = \tfrac{1}{2}}_{B}$ or $\ket{\mathbf{p},s^{1}_{p}}_{\Sigma}$ are relevant only for substantial values of velocity. Taking into account that a state produced by a Stern-Gerlach in the x direction in general must have a small velocity component $\tilde{v}^{1}$, the map shows that these discrepancies are expected to be smaller than $5\%$ for a vast majority of situations.

\section{\label{sec:4} Concluding remarks}

The results presented show that the discrepancies between the predictions obtained using Wigner's spin and the intrinsic spin are relevant, especially in the scenario where states with spin perpendicular to the measurement magnetic field are considered. However, if we take into account that the velocity of the particle in the direction parallel to the measurement field must be small, such discrepancies become less relevant, requiring considerable relativistic contributions from the other velocity components for discrepancies to be observed. In this scenario, which corresponds to a natural limitation in the design of Stern-Gerlach type experiments, discrepancies with respect to the Wigner spin become less relevant.

It is important to emphasize that the use of intrinsic spin does not invalidate the use of the Wigner basis as a way to categorize particles, since it represents a convenient basis for this purpose. However, it is important to point out that this basis does not contain the expected physical meaning for the spin, i.e., the sense of intrinsic angular momentum.

As far as relativistic quantum information theory is concerned, it is worth noting that the use of states given by the superposition of momenta will mix the contributions coming from different velocity values.  However, the conclusions presented here regarding the effect of the velocity components over the discrepancies with respect to the Wigner spin will remain valid.

Finally, we hope that the results presented here may motivate new theoretical and experimental studies concerning the relativistic spin concept.

\vspace{\baselineskip}
\begin{acknowledgments}
We are thankful for the support provided by Brazilian agencies CAPES (PROCAD2013), CNPq (\#459339/2014-1, \#312723/2018-0), FAPEG (PRONEX \#201710267000503, PRONEM \#201710267000540) and the Instituto Nacional de Ciência e Tecnologia - Informação Quântica (INCT-IQ).
\end{acknowledgments}

\appendix


\bibliography{bib/Bibliografia,bib/bibtiqr,bib/Livros}

\begin{thebibliography}{30}%
\makeatletter
\providecommand \@ifxundefined [1]{%
 \@ifx{#1\undefined}
}%
\providecommand \@ifnum [1]{%
 \ifnum #1\expandafter \@firstoftwo
 \else \expandafter \@secondoftwo
 \fi
}%
\providecommand \@ifx [1]{%
 \ifx #1\expandafter \@firstoftwo
 \else \expandafter \@secondoftwo
 \fi
}%
\providecommand \natexlab [1]{#1}%
\providecommand \enquote  [1]{``#1''}%
\providecommand \bibnamefont  [1]{#1}%
\providecommand \bibfnamefont [1]{#1}%
\providecommand \citenamefont [1]{#1}%
\providecommand \href@noop [0]{\@secondoftwo}%
\providecommand \href [0]{\begingroup \@sanitize@url \@href}%
\providecommand \@href[1]{\@@startlink{#1}\@@href}%
\providecommand \@@href[1]{\endgroup#1\@@endlink}%
\providecommand \@sanitize@url [0]{\catcode `\\12\catcode `\$12\catcode
  `\&12\catcode `\#12\catcode `\^12\catcode `\_12\catcode `\%12\relax}%
\providecommand \@@startlink[1]{}%
\providecommand \@@endlink[0]{}%
\providecommand \url  [0]{\begingroup\@sanitize@url \@url }%
\providecommand \@url [1]{\endgroup\@href {#1}{\urlprefix }}%
\providecommand \urlprefix  [0]{URL }%
\providecommand \Eprint [0]{\href }%
\providecommand \doibase [0]{https://doi.org/}%
\providecommand \selectlanguage [0]{\@gobble}%
\providecommand \bibinfo  [0]{\@secondoftwo}%
\providecommand \bibfield  [0]{\@secondoftwo}%
\providecommand \translation [1]{[#1]}%
\providecommand \BibitemOpen [0]{}%
\providecommand \bibitemStop [0]{}%
\providecommand \bibitemNoStop [0]{.\EOS\space}%
\providecommand \EOS [0]{\spacefactor3000\relax}%
\providecommand \BibitemShut  [1]{\csname bibitem#1\endcsname}%
\let\auto@bib@innerbib\@empty
\bibitem [{\citenamefont {Frenkel}(1926)}]{Frenkel1926}%
  \BibitemOpen
  \bibfield  {author} {\bibinfo {author} {\bibfnamefont {J.}~\bibnamefont
  {Frenkel}},\ }\href@noop {} {\bibfield  {journal} {\bibinfo  {journal} {Z.
  Phys.}\ }\textbf {\bibinfo {volume} {37}},\ \bibinfo {pages} {243} (\bibinfo
  {year} {1926})}\BibitemShut {NoStop}%
\bibitem [{\citenamefont {Thomas}(1926)}]{Thomas1926}%
  \BibitemOpen
  \bibfield  {author} {\bibinfo {author} {\bibfnamefont {L.~H.}\ \bibnamefont
  {Thomas}},\ }\href {https://doi.org/10.1038/117514a0} {\bibfield  {journal}
  {\bibinfo  {journal} {Nature}\ }\textbf {\bibinfo {volume} {117}},\ \bibinfo
  {pages} {514} (\bibinfo {year} {1926})}\BibitemShut {NoStop}%
\bibitem [{\citenamefont {Thomas}(1927)}]{Thomas1927}%
  \BibitemOpen
  \bibfield  {author} {\bibinfo {author} {\bibfnamefont {L.~H.}\ \bibnamefont
  {Thomas}},\ }\href {https://doi.org/10.1080/14786440108564170} {\bibfield
  {journal} {\bibinfo  {journal} {The London, Edinburgh, and Dublin
  Philosophical Magazine and Journal of Science}\ }\textbf {\bibinfo {volume}
  {3}},\ \bibinfo {pages} {1} (\bibinfo {year} {1927})},\ \Eprint
  {https://arxiv.org/abs/https://doi.org/10.1080/14786440108564170}
  {https://doi.org/10.1080/14786440108564170} \BibitemShut {NoStop}%
\bibitem [{\citenamefont {Saldanha}\ and\ \citenamefont
  {Vedral}(2012{\natexlab{a}})}]{Saldanha2012}%
  \BibitemOpen
  \bibfield  {author} {\bibinfo {author} {\bibfnamefont {P.~L.}\ \bibnamefont
  {Saldanha}}\ and\ \bibinfo {author} {\bibfnamefont {V.}~\bibnamefont
  {Vedral}},\ }\href@noop {} {\bibfield  {journal} {\bibinfo  {journal} {Phys.
  Rev. A}\ }\textbf {\bibinfo {volume} {85}},\ \bibinfo {pages} {062101}
  (\bibinfo {year} {2012}{\natexlab{a}})}\BibitemShut {NoStop}%
\bibitem [{\citenamefont {Saldanha}\ and\ \citenamefont
  {Vedral}(2012{\natexlab{b}})}]{Saldanha2012a}%
  \BibitemOpen
  \bibfield  {author} {\bibinfo {author} {\bibfnamefont {P.~L.}\ \bibnamefont
  {Saldanha}}\ and\ \bibinfo {author} {\bibfnamefont {V.}~\bibnamefont
  {Vedral}},\ }\href@noop {} {\bibfield  {journal} {\bibinfo  {journal} {New J.
  Phys.}\ }\textbf {\bibinfo {volume} {14}},\ \bibinfo {pages} {023041}
  (\bibinfo {year} {2012}{\natexlab{b}})}\BibitemShut {NoStop}%
\bibitem [{\citenamefont {Palmer}\ \emph {et~al.}(2013)\citenamefont {Palmer},
  \citenamefont {Takahashi},\ and\ \citenamefont {Westman}}]{Palmer2013}%
  \BibitemOpen
  \bibfield  {author} {\bibinfo {author} {\bibfnamefont {M.~C.}\ \bibnamefont
  {Palmer}}, \bibinfo {author} {\bibfnamefont {M.}~\bibnamefont {Takahashi}},\
  and\ \bibinfo {author} {\bibfnamefont {H.~F.}\ \bibnamefont {Westman}},\
  }\href@noop {} {\bibfield  {journal} {\bibinfo  {journal} {Ann. Phys.}\
  }\textbf {\bibinfo {volume} {336}},\ \bibinfo {pages} {505} (\bibinfo {year}
  {2013})}\BibitemShut {NoStop}%
\bibitem [{\citenamefont {Taillebois}\ and\ \citenamefont
  {Avelar}(2013)}]{Taillebois2013}%
  \BibitemOpen
  \bibfield  {author} {\bibinfo {author} {\bibfnamefont {E.~R.~F.}\
  \bibnamefont {Taillebois}}\ and\ \bibinfo {author} {\bibfnamefont {A.~T.}\
  \bibnamefont {Avelar}},\ }\href {https://doi.org/10.1103/PhysRevA.88.060302}
  {\bibfield  {journal} {\bibinfo  {journal} {Phys. Rev. A}\ }\textbf {\bibinfo
  {volume} {88}},\ \bibinfo {pages} {060302(R)} (\bibinfo {year}
  {2013})}\BibitemShut {NoStop}%
\bibitem [{\citenamefont {Caban}\ \emph
  {et~al.}(2013{\natexlab{a}})\citenamefont {Caban}, \citenamefont
  {Rembieli{\'n}ski},\ and\ \citenamefont {Wlodarczyk}}]{Caban2013}%
  \BibitemOpen
  \bibfield  {author} {\bibinfo {author} {\bibfnamefont {P.}~\bibnamefont
  {Caban}}, \bibinfo {author} {\bibfnamefont {J.}~\bibnamefont
  {Rembieli{\'n}ski}},\ and\ \bibinfo {author} {\bibfnamefont {M.}~\bibnamefont
  {Wlodarczyk}},\ }\href {https://doi.org/10.1016/j.aop.2012.12.001} {\bibfield
   {journal} {\bibinfo  {journal} {Ann. Phys.}\ }\textbf {\bibinfo {volume}
  {330}},\ \bibinfo {pages} {263} (\bibinfo {year}
  {2013}{\natexlab{a}})}\BibitemShut {NoStop}%
\bibitem [{\citenamefont {Caban}\ \emph
  {et~al.}(2013{\natexlab{b}})\citenamefont {Caban}, \citenamefont
  {Rembieli\'{n}ski},\ and\ \citenamefont {Wlodarczyk}}]{Caban2013pra}%
  \BibitemOpen
  \bibfield  {author} {\bibinfo {author} {\bibfnamefont {P.}~\bibnamefont
  {Caban}}, \bibinfo {author} {\bibfnamefont {J.}~\bibnamefont
  {Rembieli\'{n}ski}},\ and\ \bibinfo {author} {\bibfnamefont {M.}~\bibnamefont
  {Wlodarczyk}},\ }\href@noop {} {\bibfield  {journal} {\bibinfo  {journal}
  {Phys. Rev. A}\ }\textbf {\bibinfo {volume} {88}},\ \bibinfo {pages} {022119}
  (\bibinfo {year} {2013}{\natexlab{b}})}\BibitemShut {NoStop}%
\bibitem [{\citenamefont {Giacomini}\ \emph {et~al.}(2019)\citenamefont
  {Giacomini}, \citenamefont {Castro-Ruiz},\ and\ \citenamefont
  {Brukner}}]{Giacomini2019}%
  \BibitemOpen
  \bibfield  {author} {\bibinfo {author} {\bibfnamefont {F.}~\bibnamefont
  {Giacomini}}, \bibinfo {author} {\bibfnamefont {E.}~\bibnamefont
  {Castro-Ruiz}},\ and\ \bibinfo {author} {\bibfnamefont {C.}~\bibnamefont
  {Brukner}},\ }\href@noop {} {\bibfield  {journal} {\bibinfo  {journal} {Phys.
  Rev. Lett.}\ }\textbf {\bibinfo {volume} {123}},\ \bibinfo {pages} {090404}
  (\bibinfo {year} {2019})}\BibitemShut {NoStop}%
\bibitem [{\citenamefont {{Taillebois}}\ and\ \citenamefont
  {{Avelar}}(2021)}]{Taillebois2021}%
  \BibitemOpen
  \bibfield  {author} {\bibinfo {author} {\bibfnamefont {E.~R.~F.}\
  \bibnamefont {{Taillebois}}}\ and\ \bibinfo {author} {\bibfnamefont {A.~T.}\
  \bibnamefont {{Avelar}}},\ }\href@noop {} {\bibfield  {journal} {\bibinfo
  {journal} {Phys. Lett. A}\ }\textbf {\bibinfo {volume} {392}},\ \bibinfo
  {pages} {127166} (\bibinfo {year} {2021})}\BibitemShut {NoStop}%
\bibitem [{\citenamefont {Czachor}(1997)}]{Czachor1997}%
  \BibitemOpen
  \bibfield  {author} {\bibinfo {author} {\bibfnamefont {M.}~\bibnamefont
  {Czachor}},\ }\href@noop {} {\bibfield  {journal} {\bibinfo  {journal} {Phys.
  Rev. A}\ }\textbf {\bibinfo {volume} {55}},\ \bibinfo {pages} {72} (\bibinfo
  {year} {1997})}\BibitemShut {NoStop}%
\bibitem [{\citenamefont {Peres}\ \emph {et~al.}(2002)\citenamefont {Peres},
  \citenamefont {Scudo},\ and\ \citenamefont {Terno}}]{Peres2002}%
  \BibitemOpen
  \bibfield  {author} {\bibinfo {author} {\bibfnamefont {A.}~\bibnamefont
  {Peres}}, \bibinfo {author} {\bibfnamefont {P.~F.}\ \bibnamefont {Scudo}},\
  and\ \bibinfo {author} {\bibfnamefont {D.~R.}\ \bibnamefont {Terno}},\
  }\href@noop {} {\bibfield  {journal} {\bibinfo  {journal} {Phys. Rev. Lett.}\
  }\textbf {\bibinfo {volume} {88}},\ \bibinfo {pages} {230402} (\bibinfo
  {year} {2002})}\BibitemShut {NoStop}%
\bibitem [{\citenamefont {Peres}\ and\ \citenamefont
  {Terno}(2004)}]{Peres2004}%
  \BibitemOpen
  \bibfield  {author} {\bibinfo {author} {\bibfnamefont {A.}~\bibnamefont
  {Peres}}\ and\ \bibinfo {author} {\bibfnamefont {D.~R.}\ \bibnamefont
  {Terno}},\ }\href@noop {} {\bibfield  {journal} {\bibinfo  {journal} {Rev.
  Mod. Phys.}\ }\textbf {\bibinfo {volume} {76}},\ \bibinfo {pages} {93}
  (\bibinfo {year} {2004})}\BibitemShut {NoStop}%
\bibitem [{\citenamefont {Zambianco}\ \emph {et~al.}(2019)\citenamefont
  {Zambianco}, \citenamefont {Landulfo},\ and\ \citenamefont
  {Matsas}}]{Zambianco2019}%
  \BibitemOpen
  \bibfield  {author} {\bibinfo {author} {\bibfnamefont {H.~M.}\ \bibnamefont
  {Zambianco}}, \bibinfo {author} {\bibfnamefont {A.~G.}\ \bibnamefont
  {Landulfo}},\ and\ \bibinfo {author} {\bibfnamefont {G.~E.~A.}\ \bibnamefont
  {Matsas}},\ }\href@noop {} {\bibfield  {journal} {\bibinfo  {journal} {Phys.
  Rev. A}\ }\textbf {\bibinfo {volume} {100}},\ \bibinfo {pages} {062126}
  (\bibinfo {year} {2019})}\BibitemShut {NoStop}%
\bibitem [{\citenamefont {Pryce}(1948)}]{Pryce1948}%
  \BibitemOpen
  \bibfield  {author} {\bibinfo {author} {\bibfnamefont {M.~H.~L.}\
  \bibnamefont {Pryce}},\ }\href@noop {} {\bibfield  {journal} {\bibinfo
  {journal} {Proc. R. Soc. Lond. A}\ }\textbf {\bibinfo {volume} {195}},\
  \bibinfo {pages} {62} (\bibinfo {year} {1948})}\BibitemShut {NoStop}%
\bibitem [{\citenamefont {Fleming}(1965)}]{Fleming1965a}%
  \BibitemOpen
  \bibfield  {author} {\bibinfo {author} {\bibfnamefont {G.~N.}\ \bibnamefont
  {Fleming}},\ }\href@noop {} {\bibfield  {journal} {\bibinfo  {journal} {Phys.
  Rev.}\ }\textbf {\bibinfo {volume} {137}},\ \bibinfo {pages} {B 188}
  (\bibinfo {year} {1965})}\BibitemShut {NoStop}%
\bibitem [{\citenamefont {Bauke}\ \emph {et~al.}(2014)\citenamefont {Bauke},
  \citenamefont {Ahrens}, \citenamefont {Keitel},\ and\ \citenamefont
  {Grobe}}]{Bauke2014}%
  \BibitemOpen
  \bibfield  {author} {\bibinfo {author} {\bibfnamefont {H.}~\bibnamefont
  {Bauke}}, \bibinfo {author} {\bibfnamefont {S.}~\bibnamefont {Ahrens}},
  \bibinfo {author} {\bibfnamefont {C.~H.}\ \bibnamefont {Keitel}},\ and\
  \bibinfo {author} {\bibfnamefont {R.}~\bibnamefont {Grobe}},\ }\href
  {https://doi.org/10.1088/1367-2630/16/4/043012} {\bibfield  {journal}
  {\bibinfo  {journal} {New J. Phys.}\ }\textbf {\bibinfo {volume} {16}},\
  \bibinfo {pages} {043012} (\bibinfo {year} {2014})}\BibitemShut {NoStop}%
\bibitem [{\citenamefont {Hegerfeldt}(1974)}]{Hegerfeldt1974}%
  \BibitemOpen
  \bibfield  {author} {\bibinfo {author} {\bibfnamefont {G.~C.}\ \bibnamefont
  {Hegerfeldt}},\ }\href@noop {} {\bibfield  {journal} {\bibinfo  {journal}
  {Phys. Rev. D}\ }\textbf {\bibinfo {volume} {10}},\ \bibinfo {pages} {3320}
  (\bibinfo {year} {1974})}\BibitemShut {NoStop}%
\bibitem [{\citenamefont {Hegerfeldt}\ and\ \citenamefont
  {Ruijsenaars}(1980)}]{Hegerfeldt1980}%
  \BibitemOpen
  \bibfield  {author} {\bibinfo {author} {\bibfnamefont {G.~C.}\ \bibnamefont
  {Hegerfeldt}}\ and\ \bibinfo {author} {\bibfnamefont {S.~N.~M.}\ \bibnamefont
  {Ruijsenaars}},\ }\href@noop {} {\bibfield  {journal} {\bibinfo  {journal}
  {Phys. Rev. D}\ }\textbf {\bibinfo {volume} {22}},\ \bibinfo {pages} {377}
  (\bibinfo {year} {1980})}\BibitemShut {NoStop}%
\bibitem [{\citenamefont {Hegerfeldt}(1985)}]{Hegerfeldt1985}%
  \BibitemOpen
  \bibfield  {author} {\bibinfo {author} {\bibfnamefont {G.~C.}\ \bibnamefont
  {Hegerfeldt}},\ }\href@noop {} {\bibfield  {journal} {\bibinfo  {journal}
  {Phys. Rev. Lett.}\ }\textbf {\bibinfo {volume} {54}},\ \bibinfo {pages}
  {2395} (\bibinfo {year} {1985})}\BibitemShut {NoStop}%
\bibitem [{\citenamefont {Palmer}\ \emph {et~al.}(2012)\citenamefont {Palmer},
  \citenamefont {Takahashi},\ and\ \citenamefont {Westman}}]{Palmer2012}%
  \BibitemOpen
  \bibfield  {author} {\bibinfo {author} {\bibfnamefont {M.~C.}\ \bibnamefont
  {Palmer}}, \bibinfo {author} {\bibfnamefont {M.}~\bibnamefont {Takahashi}},\
  and\ \bibinfo {author} {\bibfnamefont {H.~F.}\ \bibnamefont {Westman}},\
  }\href@noop {} {\bibfield  {journal} {\bibinfo  {journal} {Ann. Phys.}\
  }\textbf {\bibinfo {volume} {327}},\ \bibinfo {pages} {1078} (\bibinfo {year}
  {2012})}\BibitemShut {NoStop}%
\bibitem [{\citenamefont {Caban}\ \emph {et~al.}(2014)\citenamefont {Caban},
  \citenamefont {Rembieli{\'n}ski}, \citenamefont {Rybka}, \citenamefont
  {Smoli{\'n}ski},\ and\ \citenamefont {Witas}}]{Caban2014}%
  \BibitemOpen
  \bibfield  {author} {\bibinfo {author} {\bibfnamefont {P.}~\bibnamefont
  {Caban}}, \bibinfo {author} {\bibfnamefont {J.}~\bibnamefont
  {Rembieli{\'n}ski}}, \bibinfo {author} {\bibfnamefont {P.}~\bibnamefont
  {Rybka}}, \bibinfo {author} {\bibfnamefont {K.~A.}\ \bibnamefont
  {Smoli{\'n}ski}},\ and\ \bibinfo {author} {\bibfnamefont {P.}~\bibnamefont
  {Witas}},\ }\href@noop {} {\bibfield  {journal} {\bibinfo  {journal} {Phys.
  Rev. A}\ }\textbf {\bibinfo {volume} {89}},\ \bibinfo {pages} {032107}
  (\bibinfo {year} {2014})}\BibitemShut {NoStop}%
\bibitem [{\citenamefont {C{\'e}leri}\ \emph {et~al.}(2016)\citenamefont
  {C{\'e}leri}, \citenamefont {Kiosses},\ and\ \citenamefont
  {Terno}}]{Celeri2016}%
  \BibitemOpen
  \bibfield  {author} {\bibinfo {author} {\bibfnamefont {L.~C.}\ \bibnamefont
  {C{\'e}leri}}, \bibinfo {author} {\bibfnamefont {V.}~\bibnamefont
  {Kiosses}},\ and\ \bibinfo {author} {\bibfnamefont {D.~R.}\ \bibnamefont
  {Terno}},\ }\href@noop {} {\bibfield  {journal} {\bibinfo  {journal} {Phys.
  Rev. A}\ }\textbf {\bibinfo {volume} {94}},\ \bibinfo {pages} {062115}
  (\bibinfo {year} {2016})}\BibitemShut {NoStop}%
\bibitem [{\citenamefont {Silva}\ \emph {et~al.}(2019)\citenamefont {Silva},
  \citenamefont {Taillebois}, \citenamefont {Gomes}, \citenamefont {Walborn},\
  and\ \citenamefont {Avelar}}]{Silva2019}%
  \BibitemOpen
  \bibfield  {author} {\bibinfo {author} {\bibfnamefont {T.~L.}\ \bibnamefont
  {Silva}}, \bibinfo {author} {\bibfnamefont {E.~R.~F.}\ \bibnamefont
  {Taillebois}}, \bibinfo {author} {\bibfnamefont {R.~M.}\ \bibnamefont
  {Gomes}}, \bibinfo {author} {\bibfnamefont {S.~P.}\ \bibnamefont {Walborn}},\
  and\ \bibinfo {author} {\bibfnamefont {A.~T.}\ \bibnamefont {Avelar}},\
  }\href@noop {} {\bibfield  {journal} {\bibinfo  {journal} {Phys. Rev. A}\
  }\textbf {\bibinfo {volume} {99}},\ \bibinfo {pages} {022332} (\bibinfo
  {year} {2019})}\BibitemShut {NoStop}%
\bibitem [{\citenamefont {Taillebois}\ and\ \citenamefont
  {Avelar}(2021)}]{Taillebois2021a}%
  \BibitemOpen
  \bibfield  {author} {\bibinfo {author} {\bibfnamefont {E.~R.~F.}\
  \bibnamefont {Taillebois}}\ and\ \bibinfo {author} {\bibfnamefont {A.~T.}\
  \bibnamefont {Avelar}},\ }\href@noop {} {\bibfield  {journal} {\bibinfo
  {journal} {Phys. Rev. A}\ }\textbf {\bibinfo {volume} {103}},\ \bibinfo
  {pages} {062223} (\bibinfo {year} {2021})}\BibitemShut {NoStop}%
\bibitem [{\citenamefont {Penrose}\ and\ \citenamefont
  {Rindler}(1984)}]{Penrose1984}%
  \BibitemOpen
  \bibfield  {author} {\bibinfo {author} {\bibfnamefont {R.}~\bibnamefont
  {Penrose}}\ and\ \bibinfo {author} {\bibfnamefont {W.}~\bibnamefont
  {Rindler}},\ }\href {https://doi.org/10.1017/CBO9780511564048} {\emph
  {\bibinfo {title} {Spinors and Space-Time}}},\ \bibinfo {series} {Cambridge
  Monographs on Mathematical Physics}, Vol.~\bibinfo {volume} {1}\ (\bibinfo
  {publisher} {Cambridge University Press},\ \bibinfo {year}
  {1984})\BibitemShut {NoStop}%
\bibitem [{\citenamefont {Wigner}(1939)}]{Wigner1939}%
  \BibitemOpen
  \bibfield  {author} {\bibinfo {author} {\bibfnamefont {E.}~\bibnamefont
  {Wigner}},\ }\href@noop {} {\bibfield  {journal} {\bibinfo  {journal} {Ann.
  Math.}\ }\textbf {\bibinfo {volume} {40}},\ \bibinfo {pages} {149} (\bibinfo
  {year} {1939})}\BibitemShut {NoStop}%
\bibitem [{\citenamefont {Tung}(1985)}]{Tung1985}%
  \BibitemOpen
  \bibfield  {author} {\bibinfo {author} {\bibfnamefont {W.-K.}\ \bibnamefont
  {Tung}},\ }\href@noop {} {\emph {\bibinfo {title} {{Group Theory in Physics -
  {A}n Introduction}}}}\ (\bibinfo  {publisher} {World Scientific Publishing
  Co. Pte. Ltd.},\ \bibinfo {year} {1985})\BibitemShut {NoStop}%
\bibitem [{\citenamefont {Caban}\ and\ \citenamefont
  {Rembieli{\'n}ski}(2005)}]{Caban2005}%
  \BibitemOpen
  \bibfield  {author} {\bibinfo {author} {\bibfnamefont {P.}~\bibnamefont
  {Caban}}\ and\ \bibinfo {author} {\bibfnamefont {J.}~\bibnamefont
  {Rembieli{\'n}ski}},\ }\href@noop {} {\bibfield  {journal} {\bibinfo
  {journal} {Phys. Rev. A}\ }\textbf {\bibinfo {volume} {72}},\ \bibinfo
  {pages} {012103} (\bibinfo {year} {2005})}\BibitemShut {NoStop}%
\end{thebibliography}%

\end{document}